\documentstyle[12pt]{article}
\begin{document}
\title{ Inflation and Transition to a Slowly Accelerating Phase 
from S.S.B. of Scale Invariance \protect\\  } \author{E.I. Guendelman and O.Katz  
\\{\it Physics Department, Ben-Gurion University, Beer-Sheva
84105, Israel}}

\maketitle
\bigskip

\begin{abstract}
 We consider the effects of adding a scale invariant $R^{2}$ term to the 
action of the scale invariant model (SIM) studied previously by one of us 
(E.I.G., Mod. Phys. Lett. A14, 1043 (1999)). The SIM belongs to the general 
class of  theories, where an integration measure independent of the metric 
is introduced. To implement scale invariance (S.I.), a dilaton  field is 
introduced. The integration of the equations of motion associated with the 
new measure gives rise to the spontaneous symmetry breaking (S.S.B) of S.I.. 
After S.S.B. of S.I. in the model with the $R^{2}$ term, it is found that 
a non trivial potential for the dilaton is generated. This potential contains
two flat regions: one associated with the Planck scale and with an 
inflationary phase, while the other flat region is associated to a very small
vacuum energy (V.E.) and is associated to the present slowly accelerated 
phase of the universe (S.A.PH). The smallness of the V.E. in the S.A.PH. 
is understood through the see saw mechanism introduced in S.I.M. 
 
\end{abstract}

\section{Introduction}
The vacuum energy density of the universe has been invoked as a 
fundamental component of the energy density of the universe at 
least in two instances. First, it has been used as the source of 
a possible inflationary phase of the early universe $^1$, which provides 
an attractive scenario for solving
 some of the fundamental puzzles of the standard Big Bang model,
like the horizon and the flatness problems as well as providing a
framework for sensible calculations of primordial density perturbations.
In recent years, with the discovery of the acceleration of the present 
universe$^2$ it appears plausible that a small vacuum energy is also 
present even today.

These two vacuum energy densities, the one of inflation and that of the 
Universe now have however a totally different scale. One wonders then
how cosmological evolution may naturally interpolate between such two 
apparently quite disconnected physical situations.

The possibility of continously connecting an inflationary phase to a slowly 
accelerated universe through the evolution of a single scalar field has been 
studied $^3$ . In this paper we propose a theoretical framework 
where such scenario is realized in a natural way.

We work in the context of a two measures theory (TMT) $^{4,5,6,7}$
and more specifically in the context of the scale invariant 
realization of such theories $^{5,6,7}$ and will show how the stated goals 
can be achieved in this context.

The paper will be organized as follows: In section 2 we review the principles
of the TMT and in particular the model studied in Ref. 5, which has 
global scale invariance. Such model gives rise, in the effective 
Einstein frame, to an effective potential for a dilaton field (needed to
implement an interesting model with global scale invariance) which has
a flat region . In section 3 we review the "cosmological see saw mechanism"
introduced in Ref. 5 and which allows to obtain a small value for
the vacuum energy without adjusting small numbers. In section 4, we 
generalize this model by adding a curvature square or simply "$R^{2}$ term" 
and show that the resulting model contains now two flat 
regions which can be of very different scales: one must be associated to the 
Planck scale and is associated to an inflationary phase while the other
flat region can be very small due to a "cosmological see-saw" mechanism.
This very small vacuum energy may be then associated to the present 
accelerated universe. The existence of two flat regions for the potential
is shown to be consequence of the s.s.b. of the scale symmetry. 
 We end with a discussion and conclusions section.

\section{The Two Measures Theories and a simple scale invariant realization}

	When studying generally covariant Lagrangian formulations of
gravitational theories, we usually consider the form
\begin{equation}
S_{1} = \int{L}\sqrt{-g} d^{4}x, g =  det g_{\mu\nu}
\end{equation}

	As it is well known, $d^{4}x$ is not a scalar but the combination
$\sqrt{-g} d^{4} x$ is a scalar. Inserting $\sqrt{-g}$,
which has the transformation properties of a
density, produces a scalar action (1), provided L is a scalar.

	One could use nevertheless other objects instead of
$\sqrt{-g}$, provided
they have the same transformation properties and achieve in this way a
different generally covariant formulation.

	For example, given 4-scalars $\varphi_{a}$ (a = 
1,2,3,4), one can construct the density
\begin{equation} 
\Phi =  \varepsilon^{\mu\nu\alpha\beta}  \varepsilon_{abcd}
\partial_{\mu} \varphi_{a} \partial_{\nu} \varphi_{b} \partial_{\alpha}
\varphi_{c} \partial_{\beta} \varphi_{d}  
\end{equation}
and consider instead of (1) the action
\begin{equation}
S_{2} =  \int L \Phi d^{4} x	
\end{equation}
L is again some scalar, which may contain the curvature (i.e. the
gravitational contribution) and a matter contribution, as is standard also
in (1).

	In the action (3) the measure carries degrees of freedom
independent of that of the metric and that of the matter fields. The most
natural and successful formulation of the theory is achieved when the
connection is also treated as an independent degree of freedom. This is
what is usually referred to as the first order formalism.

	One can notice that $\Phi$ is the total derivative of something,
for
example, one can write$^{4,5,6,7}$
\begin{equation}
\Phi = \partial_{\mu} ( \varepsilon^{\mu\nu\alpha\beta}
\varepsilon_{abcd} \varphi_{a}
  \partial_{\nu} \varphi_{b}
\partial_{\alpha}
\varphi_{c} \partial_{\beta} \varphi_{d}). 
\end{equation}

	This means that a shift of the form
\begin{equation}	
		L \rightarrow  L  +  constant	
\end{equation}
just adds the integral of a total divergence to the action (3) and it does
not affect therefore the equations of motion of the theory. The same
shift, acting on (1) produces an additional term which gives rise to a
cosmological constant.

	One can generalize this structure and allow both geometrical
objects to enter the theory and consider
\begin{equation}
S_{3} = \int L_{1} \Phi  d^{4} x  +  \int L_{2} \sqrt{-g}d^{4}x		
\end{equation}

	Now instead of  (5),  the shift symmetry can be applied only on
$L_{1}$ 
($L_{1} \rightarrow L_{1}$ + constant). Here $L_{1}$ and
$L_{2}$ are
$\varphi_{a}$  independent.

	There is a good reason not to consider mixing of  $\Phi$ and
$\sqrt{-g}$ , like
for example using
\begin{equation}
\frac{\Phi^{2}}{\sqrt{-g}} 
\end{equation}		 	

	This is because (6) is invariant (up to the integral of a total
divergence) under the infinite dimensional symmetry
\begin{equation}
\varphi_{a} \rightarrow \varphi_{a}  +  f_{a} (L_{1})	
\end{equation}
where $f_{a} (L_{1})$ is an arbitrary function of $L_{1}$ if $L_{1}$ and
$L_{2}$ are $\varphi_{a}$
independent. Such symmetry (up to the integral of a total divergence) is
absent if mixed terms (like (7)) are present.  Therefore (6) is considered
for the case when no dependence on the measure fields (MF) appears in
$L_{1}$ or $L_{2}$.

	We will study now the dynamics of a scalar field $\phi$ interacting
with gravity as given by the following action$^5$

\begin{equation}
S_{L} =  \int L_{1} \Phi d^{4} x  +  \int L_{2} \sqrt{-g}   d^{4} x
\end{equation}
\begin{equation}
L_{1} = \frac{-1}{\kappa} R(\Gamma, g) + \frac{1}{2} g^{\mu\nu}
\partial_{\mu} \phi \partial_{\nu} \phi - V(\phi) 
\end{equation}	
\begin{equation}
L_{2} = U(\phi)
\end{equation}
\begin{equation}	
R(\Gamma,g) =  g^{\mu\nu}  R_{\mu\nu} (\Gamma) , R_{\mu\nu}
(\Gamma) = R^{\lambda}_{\mu\nu\lambda}
\end{equation}
\begin{equation}
R^{\lambda}_{\mu\nu\sigma} (\Gamma) = \Gamma^{\lambda}_
{\mu\nu,\sigma} - \Gamma^{\lambda}_{\mu\sigma,\nu} +
\Gamma^{\lambda}_{\alpha\sigma}  \Gamma^{\alpha}_{\mu\nu} -
\Gamma^{\lambda}_{\alpha\nu} \Gamma^{\alpha}_{\mu\sigma}.	 
\end{equation}

The suffix $L$ in $S_{L}$ is to emphasize that here the curvature appears
only linearly.

	In the variational principle $\Gamma^{\lambda}_{\mu\nu},
g_{\mu\nu}$, the measure fields scalars
$\varphi_{a}$ and the "matter" - scalar field $\phi$ are all to be treated
as independent
variables although the variational principle may result in equations that
allow us to solve some of these variables in terms of others.

	There is the interesting possibility of implementing global 
scale invariance in such a model. Indeed, if we perform the global 
scale transformation ($\theta$ =
constant) 
\begin{equation}
g_{\mu\nu}  \rightarrow   e^{\theta}  g_{\mu\nu}	
\end{equation}
then (9) is invariant provided  $V(\phi)$ and $U(\phi)$ are of the
form  
\begin{equation}
V(\phi) = f_{1}  e^{\alpha\phi},  U(\phi) =  f_{2}
e^{2\alpha\phi}
\end{equation}
and $\varphi_{a}$ is transformed according to
\begin{equation}
\varphi_{a}   \rightarrow   \lambda_{a} \varphi_{a}  
\end{equation}
(no sum on a) which means
\begin{equation}
\Phi \rightarrow \biggl(\prod_{a} {\lambda}_{a}\biggr) \Phi \\ \equiv \lambda 
\Phi	 \end{equation}
such that
\begin{equation} 
\lambda = e^{\theta}
\end{equation}	
and	 
\begin{equation}
\phi \rightarrow \phi - \frac{\theta}{\alpha}.     	
\end{equation}

	We will now work out the equations of motion for arbitrary choice
of $V(\phi)$ and $U(\phi)$. We study afterwards the choice (15) which
allows us to
obtain the results for the scale invariant case and also to see what
differentiates this from the choice of arbitrary $U(\phi)$ and  $V(\phi)$ 
in a very
special way.

	Let us begin by considering the equations which are obtained from
the variation of the fields that appear in the measure, i.e. the
$\varphi_{a}$
fields. We obtain then  
\begin{equation}		
A^{\mu}_{a} \partial_{\mu} L_{1} = 0   	
\end{equation}
where  $A^{\mu}_{a} = \varepsilon^{\mu\nu\alpha\beta}
\varepsilon_{abcd} \partial_{\nu} \varphi_{b} \partial_{\alpha}
\varphi_{c} \partial_{\beta} \varphi_{d}$. Since it is easy to
check that  $A^{\mu}_{a} \partial_{\mu} \varphi_{a^{\prime}} =
\frac{\delta aa^{\prime}}{4} \Phi$, it follows that 
det $(A^{\mu}_{a}) =\frac{4^{-4}}{4!} \Phi^{3} \neq 0$ if $\Phi\neq 0$.
Therefore if $\Phi\neq 0$ we obtain that $\partial_{\mu} L_{1} = 0$,
 or that
\begin{equation}
L_{1} = \frac{-1}{\kappa} R(\Gamma,g) + \frac{1}{2} g^{\mu\nu}
\partial_{\mu} \phi \partial_{\nu} \phi - V = M	 
\end{equation}
where M is constant.

	Let us study now the equations obtained from the variation of the
connections $\Gamma^{\lambda}_{\mu\nu}$.  We obtain then
\begin{equation}
-\Gamma^{\lambda}_{\mu\nu} -\Gamma^{\alpha}_{\beta\mu}
g^{\beta\lambda} g_{\alpha\nu}  + \delta^{\lambda}_{\nu}
\Gamma^{\alpha}_{\mu\alpha} + \delta^{\lambda}_{\mu}
g^{\alpha\beta} \Gamma^{\gamma}_{\alpha\beta}
g_{\gamma\nu}\\ - g_{\alpha\nu} \partial_{\mu} g^{\alpha\lambda}
+ \delta^{\lambda}_{\mu} g_{\alpha\nu} \partial_{\beta}
g^{\alpha\beta}
 - \delta^{\lambda}_{\nu} \frac{\Phi,_\mu}{\Phi}
+ \delta^{\lambda}_{\mu} \frac{\Phi,_           \nu}{\Phi} =  0	
\end{equation}
If we define $\Sigma^{\lambda}_{\mu\nu}$    as
$\Sigma^{\lambda}_{\mu\nu} =
\Gamma^{\lambda}_{\mu\nu} -\{^{\lambda}_{\mu\nu}\}$
where $\{^{\lambda}_{\mu\nu}\}$   is the Christoffel symbol, we
obtain for $\Sigma^{\lambda}_{\mu\nu}$ the equation 
\begin{equation}
	-  \sigma, _{\lambda} g_{\mu\nu} + \sigma, _{\mu}
g_{\nu\lambda} - g_{\nu\alpha} \Sigma^{\alpha}_{\lambda\mu}
-g_{\mu\alpha} \Sigma^{\alpha}_{\nu \lambda}
+ g_{\mu\nu} \Sigma^{\alpha}_{\lambda\alpha} +
g_{\nu\lambda} g_{\alpha\mu} g^{\beta\gamma} \Sigma^{\alpha}_{\beta\gamma}
= 0 
\end{equation}		 
where  $\sigma = ln \chi, \chi = \frac{\Phi}{\sqrt{-g}}$.
      	
	The general solution of (23) is 
\begin{equation}
\Sigma^{\alpha}_{\mu\nu} = \delta^{\alpha}_{\mu}
\lambda,_{\nu} + \frac{1}{2} (\sigma,_{\mu} \delta^{\alpha}_{\nu} -
\sigma,_{\beta} g_{\mu\nu} g^{\alpha\beta})
\end{equation}
where $\lambda$ is an arbitrary function due to the $\lambda$ - symmetry
of the
curvature$^{8}$  $R^{\lambda}_{\mu\nu\alpha} (\Gamma)$,
\begin{equation}
\Gamma^{\alpha}_{\mu\nu} \rightarrow \Gamma^{\prime \alpha}_{\mu\nu}
 = \Gamma^{\alpha}_{\mu\nu} + \delta^{\alpha}_{\mu}
Z,_{\nu}
\end{equation} 
Z  being any scalar (which means $\lambda \rightarrow \lambda + Z$).
  
	If we choose the gauge $\lambda = \frac{\sigma}{2}$, we obtain
\begin{equation}
\Sigma^{\alpha}_{\mu\nu} (\sigma) = \frac{1}{2} (\delta^{\alpha}_{\mu}
\sigma,_{\nu} +
 \delta^{\alpha}_{\nu} \sigma,_{\mu} - \sigma,_{\beta}
g_{\mu\nu} g^{\alpha\beta}).
\end{equation}

	Considering now the variation with respect to $g^{\mu\nu}$, we
obtain
\begin{equation}	 	
\Phi (\frac{-1}{\kappa} R_{\mu\nu} (\Gamma) + \frac{1}{2} \phi,_{\mu}
\phi,_{\nu}) - \frac{1}{2} \sqrt{-g} U(\phi) g_{\mu\nu} = 0
\end{equation}
solving for $R = g^{\mu\nu} R_{\mu\nu} (\Gamma)$  from (27) and introducing 
in (21), we obtain
\begin{equation}
M + V(\phi) - \frac{2U(\phi)}{\chi} = 0
\end{equation}
a constraint that allows us to solve for $\chi$,
\begin{equation}
\chi = \frac{2U(\phi)}{M+V(\phi)}.
\end{equation}

	To get the physical content of the theory, it is convenient to go
to the Einstein conformal frame where 
\begin{equation}
\overline{g}_{\mu\nu} = \chi g_{\mu\nu}		    
\end{equation}
and $\chi$  given by (29b). In terms of $\overline{g}_{\mu\nu}$   the non
Riemannian contribution $\Sigma^{\alpha}_{\mu\nu}$
dissappears from the equations. This is because the connection
can be written as the Christoffel symbol of the metric 
$\overline{g}_{\mu\nu}$ .
In terms of $\overline{g}_{\mu\nu}$ the equations
of motion for the metric can be written then in the Einstein
form (we define $\overline{R}_{\mu\nu} (\overline{g}_{\alpha\beta}) =$  
 usual Ricci tensor in terms of the bar metric $= R_{\mu\nu}$ 
 $\overline{R}  = \overline{g}^{\mu \nu}  \overline{R}_{\mu\nu}$ )
\begin{equation}
\overline{R}_{\mu\nu} (\overline{g}_{\alpha\beta}) - \frac{1}{2} 
\overline{g}_{\mu\nu}
\overline{R}(\overline{g}_{\alpha\beta}) = \frac{\kappa}{2} T^{eff}_{\mu\nu}
(\phi)	 	
\end{equation}
where
\begin{equation}	 
T^{eff}_{\mu\nu} (\phi) = \phi_{,\mu} \phi_{,\nu} - \frac{1}{2} \overline
{g}_{\mu\nu} \phi_{,\alpha} \phi_{,\beta} \overline{g}^{\alpha\beta}
+ \overline{g}_{\mu\nu} V_{eff} (\phi)
\end{equation}

and 	
\begin{equation}
V_{eff} (\phi) = \frac{1}{4U(\phi)}  (V+M)^{2}.
\end{equation}
	
	In terms of the metric $\overline{g}^{\alpha\beta}$ , the equation
of motion of the Scalar
field $\phi$ takes the standard General - Relativity form
\begin{equation}
\frac{1}{\sqrt{-\overline{g}}} \partial_{\mu} (\overline{g}^{\mu\nu} 
\sqrt{-\overline{g}} \partial_{\nu}
\phi) + V^{\prime}_{eff} (\phi) = 0.
\end{equation} 

	Notice that if  $V + M = 0,  V_{eff}  = 0$ and $V^{\prime}_{eff} 
= 0$ also, provided $V^{\prime}$ is finite and $U \neq 0$ there. 
This means the zero cosmological constant
state
is achieved without any sort of fine tuning. That is, independently
of whether we add to $V$ a constant piece, or whether we change
the value of $M$, as long as there is still a point
where $V+M =0$, then still $ V_{eff}  = 0$ and $V^{\prime}_{eff} = 0$
( still provided $V^{\prime}$ is finite and $U \neq 0$ there). 
This is the basic feature
that characterizes the TMT and allows it to solve the 'old'
cosmological constant problem$^{4}$ at least at the classical level.

	In what follows we will study (33) for the special case of global
scale invariance, which as we will see displays additional very special
features which makes it attractive in the context of cosmology.

	Notice that in terms of the variables $\phi$,
$\overline{g}_{\mu\nu}$, the "scale"
transformation becomes only a shift in the scalar field $\phi$, since
$\overline{g}_{\mu\nu}$ is
invariant (since $\chi \rightarrow \lambda^{-1} \chi$  and $g_{\mu\nu}
\rightarrow \lambda g_{\mu\nu}$)
\begin{equation}
\overline{g}_{\mu\nu} \rightarrow \overline{g}_{\mu\nu}, \phi \rightarrow
\phi - \frac{\theta}{\alpha}.
\end{equation}

	If $V(\phi) = f_{1} e^{\alpha\phi}$  and  $U(\phi) = f_{2}
e^{2\alpha\phi}$ as
required by scale
invariance (14), (16), (17), (18), (19), we obtain from (33)
\begin{equation}
	V_{eff}  = \frac{1}{4f_{2}}  (f_{1}  +  M e^{-\alpha\phi})^{2}	
\end{equation}

	Since we can always perform the transformation $\phi \rightarrow
- \phi$ we can
choose by convention $\alpha > 0$. We then see that as $\phi \rightarrow
\infty, V_{eff} \rightarrow \frac{f_{1}^{2}}{4f_{2}} =$ const.
providing an infinite flat region as depicted in Fig. 1. Also a minimum is achieved at zero
cosmological constant for the case $\frac{f_{1}}{M} < 0$ at the point 
\begin{equation}
\phi_{min}  =  \frac{-1}{\alpha} ln \mid\frac{f_1}{M}\mid.  	
\end{equation}

	Finally, the second derivative of the potential  $V_{eff}$  at the
minimum is 
\begin{equation}
V^{\prime\prime}_{eff} = \frac{\alpha^2}{2f_2} \mid{f_1}\mid^{2} > 0
\end{equation}
if
$f_{2} > 0$,	 	
there are many interesting issues that one can raise here. The first one
is of course the fact that a realistic scalar field potential, with
massive exitations when considering the true vacuum state, is achieved in
a way consistent with the idea  of scale
invariance.

	A peculiar feature of the potential (36), is that the absolute 
value of the constant $M$, does not affect the physics of the problem, only the 
sign will have an effect. This is because if we perform a shift 
\begin{equation}
\phi \rightarrow \phi + \Delta		
\end{equation}	
in the potential (36), this is equivalent to the change in the integration
constant  M
\begin{equation}
M \rightarrow M e^{-\alpha\Delta}.	
\end{equation}

	We see therefore that if we change  M in any way, without changing
the sign of M, the only effect this has is to shift the whole potential.
The physics of the potential remains unchanged, however. This is 
reminiscent of the dilatation invariance of the theory, which involves
only a shift in $\phi$  if $\overline{g}_{\mu\nu}$   is used (see eq. (35)
).

	This is very different from the situation for two generic
functions
$U(\phi)$ and 
$V(\phi)$ in (33). There, M appears in $V_{eff}$ as a true new parameter
that
generically changes the shape of the potential $V_{eff}$, i.e. it is
impossible
then to compensate the effect of M with just a shift. For example  M will
appear in the value of the second derivative of the potential at the
minimum, unlike what we see in eq. (38), where we see that
$V^{\prime\prime}_{eff}$ (min) is M
independent.

	In conclusion, the scale invariance of the original theory is
responsible for the non appearance (in the physics) of a certain scale,
that associated to M. However, masses do appear, since the coupling to two
different measures of $L_{1}$ and $L_{2}$ allow us to introduce two
independent
couplings  $f_{1}$ and $f_{2}$, a situation which is  unlike the
standard
formulation of globally scale invariant theories, where usually no stable
vacuum state exists.

	Notice that we have not considered all possible terms consistent
with global scale invariance. Additional terms in  $L_{2}$  of the form
$e^{\alpha\phi} R$ and 
$e^{\alpha\phi} g^{\mu\nu} \partial_{\mu}\phi
\partial_{\nu}\phi$
 are indeed consistent with the global scale invariance
(14), (16), (17), (18), (19) but they give rise to a much more complicated
theory, which has been studied in a separate publication $^{7}$. There, 
when the theory is studied in the presence of fermions, the equation
that determines the ratio between the two measures becomes non linear
and the multiple solutions of the constraint equation provide an
oportunity for explaining the family structure of fermions$^{7}$ and other 
interesting cosmological questions.

We can compare the appearance of the potential $V_{eff} (\phi)$, which has
privileged
some point depending on M (for example the minimum of the potential will
have to be at some specific point), although the theory has the
"translation invariance" (35), to the physics of solitons.

	In fact, this very much resembles the appearance of solitons in a
space-translation invariant theory: The soliton solution has to be
centered at some point, which of course is not determined by the theory.
The soliton of  course breaks the space translation invariance
spontaneously, just as the existence of the non trivial potential $V_{eff}
(\phi)$
breaks here spontaneously the translations in $\phi$ space, since $V_{eff}
(\phi)$ is
not a constant.

The constant of integration $M$ plays a very important role indeed:
any non vanishing value for this constant implements, already at the 
classical level S.S.B. of scale invariance.

\section{ The Cosmological See Saw Mechanism}
	Notice  the existence for $\phi \rightarrow \infty$,
of a flat region for
$V_{eff} (\phi)$ can be nicely described as a region where the symmetry
under
translations (35) is restored.

       One should  point out that the model discussed here gives a
potential with two physically relevant parameters : $\frac{f_1^{2}}{4f_{2}}$ , which represents
the value of $V_{eff}$ as $\phi \rightarrow \infty$ ,  i.e. the strength
of the false vacuum at
the
flat region and  $\frac{\alpha^{2}f_{1}^{2}}{2f_2}$ , representing the
mass of the excitations around the
true vacuum with zero cosmological constant (achieved here without fine
tuning).

       One can consider this model as suitable for the present universe and
the almost constant vacuum energy for for large values of the scalar field
 $f_{1}^{2}/4f_{2}$  can be the source of the acceleration now. 

	Notice that a small value of $\frac{f_{1}^{2}}{f_{2}}$   can be
achieved if we let $f_{2} >> f_{1}$. In this case
$\frac{f_{1}^{2}}{f_{2}} << f_{1}$, i.e. a very small scale for the 
the vacuum energy can be achieved by a sort of see-saw mechanism that
resembles the neutrino see saw effect ( Ref. 9).
If the scale of $f_{1}$ is the electroweak scale and the scale of $f_{2}$     
is the Planck scale, then naturally we are led to a small vacuum energy for   
the present universe of the order of  $M_{EW}^{8}/M_{Pl}^{4}$ . 
This is also of the correct order of magnitude. It is exactly
what is needed in the new approach to cosmic coincidences discussed by        
Arkani-Hamed et. al. $^{10}$ .

         One should recall that a flat region of a potential is also a 
desirable feature in the case of inflationary scenarios. So far we have 
obtained a potential with a single flat region, so it could be used for 
either either providing a source of inflation or for providing a very small
vacuum energy for the present universe but not both. In the next section 
we will see how the addition of a curvature square term remedies this 
situation.

\section{ The scale invariant TMT with an additional $R^{2}$ term}

The simple model of section 2, which contains a single flat region can do a 
good job at describing either inflation or a slowly accelerated universe.
But what about explaining both, including a possible transition between
these two epochs?. 

A simple generalization of the action $S_{L}$ will do this job. 
What one needs to do is simply consider  the 
addition of a scale invariant term of the form

\begin{equation}                
S_{R^{2}} = \epsilon  \int (g^{\mu\nu} R_{\mu\nu} (\Gamma))^{2} \sqrt{-g} d^{4} x 
\end{equation} 

The total action being then $S = S_{L} + S_{R^{2}}$.
In the first order formalism $ S_{R^{2}}$ is not only globally scale invariant
but also locally scale invariant, that is conformally invariant (recall that
in the first order formalism the connection is an independent degree of freedom
and it does not transform under a conformal transformation of the metric).

Let us see what the equations of motion tell us, now with the addition of 
$S_{R^{2}}$ to the action. First of all, since the addition has been only to
the part of the action that couples to $ \sqrt{-g}$, the equations of motion
derived from the variation of the measure fields remains unchanged. That is
eq. (21) remains valid.

The variation of the action with respect to $ g^{\mu \nu}$ gives now

\begin{equation} 
 R_{\mu\nu} (\Gamma) ( \frac{-\Phi}{\kappa} + 2 \epsilon R  \sqrt{-g}) +
\Phi \frac{1}{2} \phi,_{\mu} \phi,_{\nu} - 
\frac{1}{2}(\epsilon R^{2} + U(\phi)) \sqrt{-g} g_{\mu\nu} = 0                     
\end{equation}

It is interesting to notice that if we contract this equation with 
 $ g^{\mu \nu}$ , the $\epsilon$ terms do not contribute. This means 
that the same value for the scalar curvature $R$ is obtained as in section 2,
 if we express our result in terms of $\phi$, its derivatives and
$ g^{\mu \nu}$ .
Solving the scalar curvature from this and inserting in the other 
$\epsilon$ - independent equation $L_{1} = M$  we get still the same 
solution for the ratio of the measures which was found in section  2,
i.e. $\chi =  \frac{\Phi}{\sqrt{-g}}  = \frac{2U(\phi)}{M+V(\phi)}$.

In the presence of the $\epsilon R^{2} $ term in the action, eq. (22)
gets modified so that instead of $\Phi$,  $\Omega$  =  
$\Phi - 2 \epsilon R \sqrt{-g}$ appears. This in turn implies that
eq. (23) mantains its form but where $\sigma$ is replaced by 
$\omega  = ln (\frac{\Omega}{\sqrt{-g}}) =
 ln ( \chi -2\kappa \epsilon R)$,
where once again, 
$\chi =  \frac{\Phi}{\sqrt{-g}} = \frac{2U(\phi)}{M+V(\phi)}$.

Following then the same steps as in section 2, we can then verify that the
connection is the Christoffel symbol of the metric $\overline{g}_{\mu\nu}$
given by 

\begin{equation} 
\overline{g}_{\mu\nu}   = (\frac{\Omega}{\sqrt{-g}}) g_{\mu\nu} 
 = (\chi -2\kappa \epsilon R) g_{\mu\nu}
\end{equation}  

$\overline{g}_{\mu\nu} $ defines now the "Einstein frame". Equations (42)
can now be expressed in the "Einstein form"  

\begin{equation}                                                              
\overline{R}_{\mu\nu} -  \frac{1}{2}\overline{g}_{\mu\ \nu} \overline{R} = 
\frac{\kappa}{2} T^{eff}_{\mu\nu}             
\end{equation}                                                                

where 
 
\begin{equation}
 T^{eff}_{\mu\nu} = 
\frac{\chi}{\chi -2 \kappa \epsilon R} (\phi_{,\mu} \phi_{,\nu} - \frac{1}{2} \overline     
{g}_{\mu\nu} \phi_{,\alpha} \phi_{,\beta} \overline{g}^{\alpha\beta})          
+ \overline{g}_{\mu\nu} V_{eff} 
\end{equation}

where 

\begin{equation}
 V_{eff}  = \frac{\epsilon R^{2} + U}{(\chi -2 \kappa \epsilon R)^{2} }
\end{equation}

Here it is satisfied that $\frac{-1}{\kappa} R(\Gamma,g) + 
\frac{1}{2} g^{\mu\nu}\partial_{\mu} \phi \partial_{\nu} \phi - V = M $,
equation that expressed in terms of $ \overline{g}^{\alpha\beta}$
 becomes

$\frac{-1}{\kappa} R(\Gamma,g) + (\chi -2\kappa \epsilon R)
\frac{1}{2} \overline{g}^{\mu\nu}\partial_{\mu} \phi \partial_{\nu} \phi - V = M$.
 This allows us to solve for $R$ and we get,

\begin{equation}
R = \frac{-\kappa (V+M) +\frac{\kappa}{2} \overline{g}^{\mu\nu}\partial_{\mu} \phi \partial_{\nu} \phi \chi} 
{1 + \kappa ^{2} \epsilon \overline{g}^{\mu\nu}\partial_{\mu} \phi \partial_{\nu} \phi}  
\end{equation}

Notice that 
 if we express $R$ in 
terms of $\phi$, its derivatives and $ g^{\mu \nu}$, the result is the 
same as in section 2, this is not true anymore once we express
 $R$ in terms of $\phi$, its derivatives and $\overline{g}^{\mu\nu}$.

In any case, once we insert (47) into (46),
we see that the  effective potential (46) depends on the derivatives of the 
scalar field. It acts as a normal scalar field potential under the 
conditions of slow rolling  or low gradients and in the case the
scalar field is near the region $M+V(\phi ) = 0$.

Notice that since 
$\chi =   \frac{2U(\phi )}{M+V(\phi )}$,
then if ${M+V(\phi) = 0}$, then, as in section 2, we obtain that 
 $ V_{eff}  =  V'_{eff} =  0$ at that point without fine tuning
(here by $ V'_{eff}$ we mean the derivative  of $ V_{eff}$ with
respect to the scalar field $\phi$, as usual).

In the case of the scale invariant case, where $V$ and $U$ are given by
equation (15), it is interesting to study the shape of $ V_{eff} $ 
as a function of $\phi$
in the case of a constant $\phi$, in which case $ V_{eff} $ can be
regarded as a real scalar field potential. Then from (47) we get
$R = -\kappa (V+M)$, which inserted in (46) gives,
\begin{equation} 
 V_{eff}  =    
\frac{(f_{1} e^{ \alpha \phi }  +  M )^{2}}{4(\epsilon \kappa ^{2}(f_{1}e^{\alpha \phi}  +  M )^{2} + f_{2}e^{2 \alpha \phi })}    
\end{equation}

a typical shape for $ V_{eff}$ for $ \epsilon = 0 $ and $\frac{f_{1}}{M} < 0$
 is depicted in Fig. 1.
This corresponds to the potential of section 2. In contrast, for
 $ \epsilon > 0 $, the corresponding potential shows two flat regions as it 
is explicitly shown in Fig. 2 (for $ \frac{f_{1}}{M} <0 $) and in  Fig. 3 
(for $ \frac{f_{1}}{M} >0 $).    

The limiting values of $ V_{eff} $ are:

First, for asymptotically 
large positive values, ie. as $ \alpha\phi \rightarrow  \infty $,   
we have
$V_{eff} \rightarrow
\frac{f_{1}^{2}}{4(\epsilon \kappa ^{2} f_{1}^{2} + f_{2})} $
and under the
see saw conditions discussed in section 3, we see that the correction in the
denominator, i.e. $\epsilon \kappa ^{2} f_{1}^{2} $, is very small
compared to the other term in the denominator, $f_{2}$. This is easy to see,
since
 $ \kappa ^{2} = \frac{1}{M_{Pl}^{4}}$ and we took in section 3 that 
$f_{1} = M_{EW}^{4} << f_{2} = M_{Pl}^{4}$ , so that indeed
$\epsilon \kappa ^{2} f_{1}^{2} <<  f_{2} $ and the see saw effect
discussed in section 3 is then unchanged by the presence of an $R^{2}$
term.

Second, for asymptotically large but negative values of the scalar field,
that is as $\alpha \phi \rightarrow - \infty  $ ,  we have: 
$ V_{eff} \rightarrow \frac{1}{4\epsilon \kappa ^{2}}$ , 
which if $\epsilon $
is a number of order one, it means we have an energy density determined by the 
Planck scale. This region is suitable for inflation.

In these two asymptotic regions ($\alpha \phi \rightarrow  \infty  $ 
and $\alpha \phi \rightarrow - \infty  $) an examination of the scalar
field equation reveals that a constant scalar field configuration is a
solution of the equations, as is of course expected from the flatness of
the effective potential in these regions.

Notice that in all the above discussion it is fundamental that $ M\neq 0$.
If $M = 0$ the potential becomes just a flat one, 
$V_{eff} = \frac{f_{1}^{2}}{4(\epsilon \kappa ^{2} f_{1}^{2} + f_{2})}$
everywhere (not only at high values  of $\alpha \phi$). The other flat 
region associated to the Planck scale and inflation is lost. That is the
spontaneous generation of the constant of integration $M$ plays  
a fundamental role in the construction of the scenario where the universe
starts in an inflationary phase and ends up in a slowly accelerated phase.
As we discussed in section 2, $ M\neq 0$ implies the we are considering a 
situation with S.S.B. of scale invariance.

\section{Discussion and Conclusions}
We have seen how the addition of an $R^{2}$ term gives rise to a 
situation where it is possible to interpolate between an inflationary 
phase and a slowly accelerated phase. This is evident after reformulating 
the theory in the "Einstein frame" and considering the situation of slow rolling 
(i.e. spacetime gradients of the scalar field are small).

We find that this is accomplished because the effective potential has now two
asymptotically flat regions instead of only one found in Ref. 5 . The 
$\epsilon R^{2}$ term is shown to have the effect of cutting off the
part of the effective potential where it went to infinity. Now instead,
this region becomes  flat. The hight of this flat region is determined by
the Planck scale.

In some previous papers $^{11, 12}$ "quintessential inflation" in the context of TMT
was studied. In Ref. 11, starting from eq. (33) some special choices for
$V$ and $U$ were shown to give a "quintesential form" for the potential
which could allow an inflationary phase and then approach to a slowly
accelerated phase also. These 
choices break scale invariance and as a result there is a great degree of
arbitrariness on the way the resulting potential turns out. These problems are
not present in the formulation presented here. In Ref. 12, 
in the context of a scale invariant theory, a potential containing a flat 
region followed by a decaying to zero piece was obtained. This could also
in principle provide a scenario for inflation followed by a slowly 
accelerated universe. The problem with the model of Ref. 12 is its 
complexity, for example four index field strengths have to be introduced, 
in addition to the the field content of the model explained here. On the
other hand, only one measure ($\Phi$ and not $\sqrt{-g}$) needs to be used.

The model we discuss here belongs still to the general class of models refered 
to as 'quintessence' (see Refs.3 and 13 for some papers on this subject),
if we mean by this that the energy density of the 
universe is dictated by the dynamics of a scalar field (some people
would add to the definition the condition that the scalar potential  
goes to zero
as the univese evolve, in this case, this depends on our prejudices
concerning the history of the universe, i.e. if ends up in the
absolute minimum of fig. 2 in the very far future of the history of  
the universe,
 for example). Such quintessence scenarios, without
the additional ingredient of an early inflation were discussed in Refs.
5,6,7 as well in a totally scale invariant framework. The addition of the
 $\epsilon R^{2}$  allows a succesfull generalization that allows inflation 
to be incorporated as we have seen.

The addition of the $\epsilon R^{2}$  term does not affect the cosmological
see saw effect for the other asymptotically flat region, that is the see saw 
effect is stable under the addition of the $\epsilon R^{2}$ terms. The 
$\epsilon $  contributions are subdominant. Therefore we get in the new 
model two flat regions:  one associated to a high 
(Planck) scale and the other which is associated to a very small vacuum
energy density. This is indeed a promising 
framework for describing the full evolution of the Universe starting from 
inflation and ending up in the slowly accelerated phase today.

In order for the whole idea to work we need to take $M \neq 0$, so the
effective potential has indeed two flat regions.  $M \neq 0$ means that
we are considering a situation with S.S.B. of scale symmetry. We can
therefore say that the  S.S.B. of scale symmetry is the source of
the inflationary phase and that of a smooth transition to a slowly 
accelerated phase today.

As we have seen, in order to get two flat regions instead of just one,
it is needed that $ M\neq 0$, which corresponds to considering a situation 
where S.S.B. of scale invariance has taken place.

It should be pointed out also that the $R^{2}$ theory studied here,  
in the context 
of an action that contains a measure $\Phi$ independent of the metric 
and with the use
of the first order formalism (the connections are considered as 
independent degrees of
freedom in the action principle), leads to equations of motion that are only 
second order, i.e. only second derivatives of the metric and the
matter fields appear (although higher powers of the derivatives of the scalar
field do appear).
This after the new
measure $\Phi$ and
the connections are solved (through the equations of motion) in terms 
of the metric
and matter fields and after we express our results in the Einstein frame.

This is in contrast to the usual $R^{2}$ theories in the second order 
formalism
and with standard measure everywhere in the action$^{14}$. In this case, 
these theories
lead to fourth order equatios for the metric field. In Ref. 14 it was shown
that the fourth order structure of the equations can be reformulated as a
system of second order equations which contain however one additional degree
of freedom, a scalar field. In the case of the  usual $R^{2}$ theories this
scalar field contains a potential with a flat region, as that of Fig. 1,
with an energy scale which is of the Planck scale. A second flat region is
missing however so that the model can describe inflation but not inflation
plus a slowly accelerated phase. For studies of inflation in
 the usual $R^{2}$ theories in the second order formalism, see Ref. 15.

\section{Acknowledgements}

We would like to thank E.Bogomolny and E.Nissimov 
for much help in the preparation of this manuscript and to 
R. Brustein,  V. Burdyuzha,
 A. Davidson  A. Kaganovich and  E.Nissimov for conversations on
the subjects discussed here.

\break

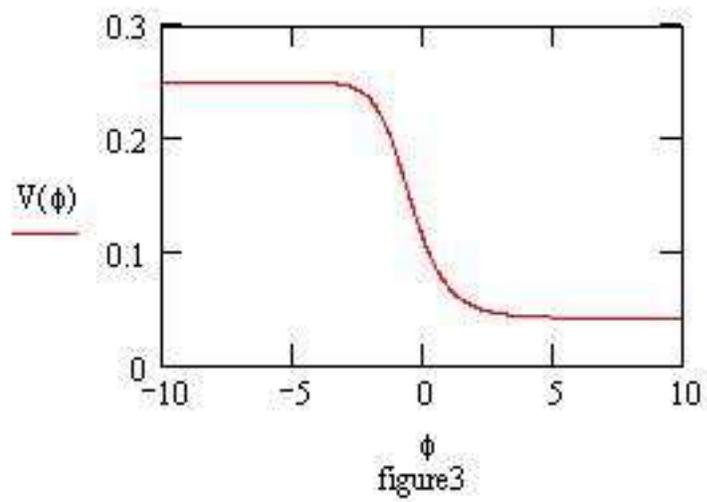
\end{document}